# A model updating procedure to enhance structural analysis in the FE code NOSA-ITACA

**Maria Girardi**

Researcher, PhD, Institute of Information Science and Technologies "A. Faedo", ISTI-CNR

via G. Moruzzi 1, I-56124 Pisa – Italy

e-mail: maria.girardi@isti.cnr.it

**Cristina Padovani**

Research Director, Dr., Institute of Information Science and Technologies "A. Faedo", ISTI-CNR

via G. Moruzzi 1, I-56124 Pisa – Italy

e-mail: cristina.padovani@isti.cnr.it

**Daniele Pellegrini**

Researcher, PhD, Institute of Information Science and Technologies "A. Faedo", ISTI-CNR

via G. Moruzzi 1, I-56124 Pisa – Italy

e-mail: daniele.pellegrini@isti.cnr.it

**Leonardo Robol**

Researcher, PhD, Institute of Information Science and Technologies "A. Faedo", ISTI-CNR

via G. Moruzzi 1, I-56124 Pisa – Italy

e-mail: leonardo.robol@isti.cnr.it

**ABSTRACT**:

This paper describes the model updating procedure implemented in NOSA-ITACA, a finite-element code for the structural analysis of masonry constructions of historical interest. The procedure, aimed at matching experimental frequencies and mode shapes, allows fine-tuning calculation of the free parameters in the model. The numerical method is briefly described, and some issues related to its robustness are addressed. The procedure is then applied to a simple case study and two historical structures in Tuscany, the Clock Tower in Lucca and the Maddalena bridge in Borgo a Mozzano.

**KEYWORDS**:

Masonry Buildings, Ambient vibration monitoring, Finite Element analysis, Model updating

**Introduction**

Finite element model updating is a procedure aimed at calibrating the finite-element (FE) model of a structure in order to match the numerical results with the experimental ones. In structural mechanics, model updating techniques are used in conjunction with vibration measurements to determine unknown system characteristics, such as material properties, constraints, etc. (Friswell et al. 1995). The resulting updated FE model can then be used to conduct more reliable structural analyses of the buildings under consideration. FE model updating is based on the solution of a constrained minimum problem, whose objective function is generally expressed as the discrepancy between experimental and numerical quantities, such as natural frequencies and mode shapes. Often, the problem is formulated as a nonlinear least square minimization.

The assessment of boundary conditions and material properties is crucial in the field of cultural heritage, where the mechanical properties of materials are rarely known accurately a priori, due to the lack of information on historical structures' construction stages and the degradation processes that they may have undergone. Application of FE model updating to ancient masonry buildings is relatively recent. Most of the papers devoted to vibration-based model updating – a detailed list is given in Girardi et al. 2018a – consider preliminary FE models that are fine-tuned using the dynamic characteristics determined through system identification techniques. These studies employ commercial codes to perform the modal analysis of the FE models, while model updating procedure is implemented separately. Many papers adopt a trial and error approach; see, for example, Costa et al. 2015, and Bayraktar et al. 2010, in which a manual fine-tuning procedure is used for FE model updating. To this end, the authors kept the mechanical properties fixed and updated the FE model of a bridge by adjusting the boundary conditions. A few recent contributions deal with automated FE model updating of a masonry bastion in Turkey (Altunişik et al. 2018) and the FE model updating of an earthquake-damaged masonry tower carried out by means of an advanced surrogate-assisted evolutionary algorithm (Bassoli et al. 2018). In Bautista-De Castro et al. 2018, the FE model updating of a concrete bridge in Portugal is conducted via a coarse to fine calibration strategy based on the Douglas-Reid and non-linear least squares approaches. Also worthy of mention is the study by Ferraioli et al. 2018, in which manual tuning of the uncertain parameters, followed by a sensitivity-based model updating technique (Bakir et al. 2007), is applied to the Santa Maria a Vico bell tower. Alternative approaches are followed in Compan et al. 2017, where a genetic algorithm is used to solve the optimization problem for the Chapel of the Würzburg Residence and in De Falco et al. 2018, where a Bayesian approach is applied to the Maddalena bridge in borgo a Mozzano and compared with the algorithm adopted in this paper.

A numerical method aimed at minimizing the discrepancy between the numerical and experimental natural frequencies has been proposed in Girardi et al. 2018a. The algorithm, based on the construction of local parametric reduced-order models embedded in a trust region scheme, has been implemented in NOSA-ITACA, a noncommercial FE code developed at the Institute of Information Science and Technologies "A. Faedo", ISTI-CNR (Binante et al. 2017). In particular, the algorithm for solving the constrained minimum problem, integrated in NOSA-ITACA, exploits the structure of the stiffness and mass matrices and the fact that only a few of the smallest eigenvalues have to be calculated. This new procedure reduces both the overall computation time of the numerical process and user effort, thus providing the scientific and technical communities with efficient algorithms specific for FE model updating.

In this paper, the method proposed in Girardi et al. 2018a and applied in Girardi et al. 2018b and De Falco et al. 2018 has been generalized in order to minimize the discrepancy between the numerical and experimental frequencies and the mode shapes as well. The generalized method is outlined, and then tested on a simple case study and on two actual structures of historical interest located in the region of Tuscany: the Clock Tower (Torre delle Ore) in Lucca and the Maddalena bridge in Borgo a Mozzano. The case studies are analyzed in detail by investigating the possibility of optimizing different parameters, as well as by performing sensitivity analyses aimed at assessing the accuracy of the recovered results. For the sake of comparison, we have ran a general purpose optimizer and show the effectiveness of the proposed method in terms of computation times.

FE model updating belongs to the class of inverse problems, as it is based on using the results of experimental measurements to infer the values of mechanical properties that characterize a structure. Within this framework, it is essential to assess the reliability of the parameters obtained through the model updating procedures. For instance, it is common practice to investigate the sensitivity of the solution to background noise. A robust method should be able to warn the user if minimal perturbation of the input (the experimental frequencies and modes) leads to large changes in the output. When this is the case, special techniques should be adopted to avoid misleading parameter identification (Mares et al. 2002). Black box methods often hide this information from the user. In the following we discuss how to evaluate the reliability of the solutions in the context of our trust-region based optimization framework.

**Model updating in NOSA-ITACA**

NOSA-ITACA (Girardi et al., 2015) is a software package for structural analysis developed by the Mechanics of Materials and Structures Laboratory of ISTI-CNR and freely available at www.nosaitaca.it. NOSA-ITACA is the result of the integration of the finite element code NOSA into the open-source graphical platform SALOME (www.salome-platform.org) and contains an extensive library of about 30 element types. The code includes modal, static and dynamic analysis in both linear and nonlinear case. In particular, the constitutive equation of no-tension materials (Lucchesi et al. 2008) is implemented to model masonry constructions, even in non-isothermal conditions (Padovani et al., 2010).

Over the last three years NOSA-ITACA has undergone several modifications and updates. In particular, special attention has been addressed to modal analysis (Porcelli et al, 2015) and model updating (Girardi et al., 2018a).

Although the masonry materials constituting historical buildings have different strengths under tension and compression, modal analysis, which is based on the assumption that these materials are linear elastic, is widely used in applications and provides important qualitative information on the dynamic behavior of such structures. Modal analysis involves solving the constrained generalized eigenvalue problem

$$\mathbf{K}\mathbf{v} = \omega^2 \mathbf{M}\mathbf{v}, \quad \text{subject to} \quad \mathbf{C}\mathbf{v} = \mathbf{0}, \tag{1}$$

with $\mathbf{C} \in \mathbb{R}^{h \times n}$, and $h \ll n$. $\mathbf{K}$ and $\mathbf{M} \in \mathbb{R}^{n \times n}$ are respectively the stiffness and mass matrices of the structure discretized into finite elements. $\mathbf{v} \in \mathbb{R}^n$ is the vector of the structure's degrees of freedom, the integer $n$ is the system's total number of degrees of freedom, which is generally very large, since it depends on the degree of discretization of the problem. The right part of equation (1) expresses the fixed constraints and the master-slave relations assigned to $\mathbf{v}$. For further details, we refer the reader to Porcelli et al. 2015 and Girardi et al. 2017.

The eigenvalues $\omega_i^2$ of (1) are linked to the natural frequencies, or eigenfrequencies $f_i$ of the structure via the relation $f_i = \omega_i / 2\pi$, and the eigenvectors $\mathbf{v}^{(i)}$ are the corresponding mode shape vectors, or eigenmodes. Together with the natural frequencies, the mode shapes furnish qualitative information on the structure's deformations under dynamic loads.

Measuring the vibrations of masonry structures is common practice for assessing their dynamic behavior and determining their natural frequencies and mode shapes. Model updating techniques are driven by vibration measurements to determine the structure's characteristics, such as the material properties (Young's modulus, Poisson's ratio, mass density), constraints, and so forth, which are generally unknown.

The model updating problem can be reformulated as an optimization problem by assuming that the stiffness and mass matrices, $\mathbf{K}$ and $\mathbf{M}$, are functions of parameter vector $\mathbf{x}$. We use the notation

$$\mathbf{K} = \mathbf{K}(\mathbf{x}), \qquad \mathbf{M} = \mathbf{M}(\mathbf{x}), \qquad \mathbf{x} \in \mathbb{R}^p \qquad (2)$$

to denote this dependency. The set of valid choices for the parameters is denoted by $\Omega$. Within this framework, we assume that the set $\Omega$ is a $p$-dimensional box, that is

$$\Omega = [a_1, b_1] \times [a_2, b_2] \times ... \times [a_p, b_p], \qquad (3)$$

for certain values $a_i < b_i$, $i = 1...p$.

By taking (2) into account, problem (1) can be rewritten as

$$\mathbf{K}(\mathbf{x})\mathbf{v}(\mathbf{x}) = \omega(\mathbf{x})^2 \mathbf{M}(\mathbf{x})\mathbf{v}(\mathbf{x}), \text{ for } \mathbf{x} \in \Omega. \qquad (4)$$

Our ultimate aim is to determine the optimal value of $\mathbf{x}$ that minimizes a certain cost functional $\Phi(\mathbf{x})$ within the box $\Omega$, that is, to solve the optimization problem

$$\min_{\mathbf{x} \in \Omega} \Phi(\mathbf{x}), \qquad (5)$$

which is a particular inverse problem, as it is aimed at improving the mathematical model of the structure under examination using measurements carried out on it.

The objective function $\Phi(\mathbf{x})$ involves the frequencies and mode shapes we wish to match, and is thus nonlinear. If we need to match $q$ frequencies and $q$ mode shapes, we choose a suitable weight vector $\mathbf{w} = (w_1, ..., w_{2q})$, with $w_i \geq 0$, and define the functional $\Phi(\mathbf{x})$ as follows:

$$\Phi(\mathbf{x}) = \sum_{i=1}^{q} w_i^2 [\hat{f}_i - f_i(\mathbf{x})]^2 + w_{q+i}^2 [1 - \gamma_i(\mathbf{x})]^2. \qquad (6)$$

$\hat{\mathbf{f}}$ is the vector of the measured frequencies, and $\mathbf{f}(\mathbf{x}) = \frac{1}{2\pi}\sqrt{(\ )}$ the vector of numerical frequencies, with $\Lambda(\mathbf{x})$ the one containing the smallest $q$ eigenvalues of (4), ordered according to their magnitude. Scalars $\gamma_i(\mathbf{x})$, given by

$$\gamma_i(\mathbf{x}) = \frac{|\mathbf{v}^{(i)}(\mathbf{x}) \cdot \hat{\mathbf{v}}^{(i)}|}{\|\mathbf{v}^{(i)}(\mathbf{x})\| \, \|\hat{\mathbf{v}}^{(i)}\|}, \quad i = 1...q, \qquad (7)$$

are the MAC (Modal Assurance Criterion) indicators (Brincker et al. 2015) and measure the correlation between the *i*th experimental mode shape, $\hat{\mathbf{v}}^{(i)}$, and the *i*th eigenvector, $\mathbf{v}^{(i)}(\mathbf{x})$ in (4), after projecting the latter on a subspace spanned by the reduced number of degrees of freedom available for the experimental mode shapes. The number *p* of parameters to be optimized is expected to be less than *q*. The vector $\mathbf{w}$ encodes the weight that should be given to each frequency and mode shape in the optimization scheme. If the goal is to minimize the distance between the vectors of the measured and computed frequencies in the usual Euclidean norm, then $w_i = 1$, $i = 1...q$, should be chosen. If, instead, relative accuracy on the frequencies is desired, $w_i = \hat{f}_i^{-1}$ is a natural choice. If some frequencies or mode shapes need to be ignored, we can set the corresponding component of $\mathbf{w}$ to zero. In order to keep the scaling uniform, the weight vector is always normalized in order to have $\|\mathbf{w}\| = 1$. For the case studies presented in the paper, experimental natural frequencies and mode shapes are identified via the Stochastic Subspace Identification (Reynders et al, 2014). As this method provides the variance of the modal properties, it allows estimating their accuracy and so choosing suitable values for the weight vector. For the eigenmodes, weights are typically fixed to 0.1. This value reflects the accuracy on the information retrieved from the identification phase in which, for our case studies, eigenvectors are obtained with one magnitude lower accuracy than the corresponding frequencies.

When the FE model is very large, it is convenient to use model reduction techniques to reduce its size to a more manageable order. If, as in equation (4), the model depends on parameters, it is not a trivial matter to obtain a reduced parametric model that accurately reflects the behavior of the original for all possible parameter values. A Lanczos-based projection strategy tailored to the needs of the FE analysis of masonry structures is presented in Girardi et al. 2018a and Girardi et al. 2018b. By modifying the projection scheme used to compute the first eigenvalues (and corresponding eigenvectors) of problem (4), we obtain local parametric reduced-order models that, embedded in a trust region scheme, are the basis for an efficient algorithm that minimizes objective function (5), starting from a point $\mathbf{x}_s$ belonging to $\Omega$. The case $w_i \geq 0, i = q+1,...,2q$, considered in this paper to take into account the mode shapes as well, can be solved by suitably generalizing the numerical procedure proposed in Girardi et al. 2018 a.

In the following, we provide a few details of the numerical procedure. We refer the reader to Girardi et al. 2018a for a more in-depth analysis of the algorithm. Trust region methods (Conn et al. 2000) require a scheme that, given any expansion point in the optimization domain, creates a local model of the objective function that is cheap to evaluate, and at least first-order accurate. To this aim, we consider the Lanczos method to compute the smallest eigenvalues and eigenvectors of interest at the expansion point. The method creates a linear subspace where the problem (4) is projected, and where accurate approximations of the eigenpairs can be recovered. A remarkable feature of the Lanczos projection is that, for symmetric definite matrices, it is possible to enlarge the subspace until a certain accuracy is guaranteed for the eigenvalues of interest, and the convergence for smaller eigenvalues is typically very fast. For instance, in the examples reported in the last section, dimensions between 20 and 40 are often enough to approximate around 5 to 10 eigenvalues to more than 10 digits for a number $n$ of degrees of freedom of the full-scale model in the order of $10^5$. We use this subspace to approximate the eigenvalues and eigenvectors in a neighborhood of the expansion point, and then build the reduced model for the trust region. The local model is used to identify descent directions by locally optimizing it with a generic local optimizer. This task has a negligible cost in the optimization procedure, being the local model very cheap to evaluate.

The model is used in the region where it is accurate enough to provide useful information on the descent directions; this can be guaranteed by suitably resizing the trust region, if necessary. It can be proven that, when the local models are accurate, convergence to a local minimizer is guaranteed. The procedure proposed is fully automatic, avoids managing the optimization problem by trial and error (as performed in a number of recent papers), and provides insight into the reliability of the solution, as described in the next subsection.

The numerical procedure is sketched in the flowchart shown in Figure 1.

*Sensitivity analysis, condition number, noise and number of parameters*

For $\mathbf{W} = diag(\mathbf{w})$, given the residual function

$$\mathbf{r}(\mathbf{x}) = \mathbf{W}\left(\hat{f}_1 - f_1(\mathbf{x}),...,\hat{f}_q - f_q(\mathbf{x}), 1-\gamma_1(\mathbf{x}),...,1-\gamma_q(\mathbf{x})\right)^T, \quad (8)$$

the objective function (6) becomes

$$\Phi(\mathbf{x}) = \|\mathbf{r}(\mathbf{x})\|^2, \quad (9)$$

and therefore the minimum problem (5) is a nonlinear least squares problem.

The gradient of $\Phi$ is

$$\nabla\Phi(\mathbf{x}) = 2\mathbf{J}(\mathbf{x})^T \mathbf{r}(\mathbf{x}), \quad (10)$$

with $\mathbf{J}(\mathbf{x})$ the Jacobian of the residual function $\mathbf{r}(\mathbf{x})$. Let us consider the singular value decomposition (SVD) of $\mathbf{J}(\mathbf{x}_0) \in \mathbb{R}^{2q \times p}$ for a fixed value of $\mathbf{x}_0$,

$$\mathbf{J}(\mathbf{x}_0) = \mathbf{Q}\mathbf{\Sigma}\mathbf{Z}^T, \quad (11)$$

where $\mathbf{Q} \in \mathbb{R}^{2q \times 2q}$ and $\mathbf{Z} \in \mathbb{R}^{p \times p}$ are unitary matrices, and the only nonzero elements of $\mathbf{\Sigma} \in \mathbb{R}^{2q \times p}$, its diagonal elements, are the singular values $\sigma_1 \geq \sigma_2 \geq ... \geq \sigma_p \geq 0$ of $\mathbf{J}(\mathbf{x})$, with $p \leq q$. Neglecting terms of order $o(\|\mathbf{x}-\mathbf{x}_0\|)$, in view of (11), we have

$$\mathbf{r}(\mathbf{x}) = \mathbf{r}(\mathbf{x}_0) + \mathbf{J}(\mathbf{x})(\mathbf{x}-\mathbf{x}_0) = \mathbf{r}(\mathbf{x}_0) + \sum_{i=1}^{p} \mathbf{q}^{(i)} \sigma_i \mathbf{z}^{(i)T}(\mathbf{x}-\mathbf{x}_0), \quad (12)$$

with $\mathbf{q}^{(i)}$ and $\mathbf{z}^{(i)}$ the column vectors of the matrices $\mathbf{Q}$ and $\mathbf{Z}$, respectively. Thus, from (12), it follows that, in a small neighborhood of $\mathbf{x}_0$, function $\mathbf{r}(\mathbf{x})$ will change the most in the direction $\mathbf{z}^{(1)}$ and the least in the direction $\mathbf{z}^{(p)}$. In particular, the change in the residual function $\mathbf{r}(\mathbf{x})$ depends linearly on the step $\mathbf{x} - \mathbf{x}_0$, and the ratio is given exactly by $\sigma_i$ if $\mathbf{x} - \mathbf{x}_0$ is aligned with $\mathbf{z}^{(i)}$.

To assess the robustness of the inverse problem (5) (i.e., its resilience to noise), we are interested in characterizing the opposite process: how much does the solution change if we perturb the input data slightly? This question is answered by resorting to the condition number.

For linear least squares problems the condition number is directly linked to the singular values of the Jacobian of the objective function (Demmel, 1997). For the nonlinear case, it is natural to consider the linearized problem (12) around a candidate minimum $\mathbf{x}_0$.

Imposing first-order optimality condition, that is asking that $\nabla \mathbf{r}(\mathbf{x}_0) = \mathbf{0}$, yields the orthogonality condition $\mathbf{J}(\mathbf{x}_0)^T \mathbf{r}(\mathbf{x}) = \mathbf{0}$. Let us write the residual (8) in the form $\mathbf{r}(\mathbf{x}) = \mathbf{a}(\mathbf{x}) - \mathbf{b}$, where $\mathbf{b}$ contains the experimental data, in our case frequencies and eigenmodes, properly weighted according to our weighting choices. Assume that the right hand side $\mathbf{b}$ is known up to an experimental uncertainty $\boldsymbol{\delta b}$. In view of (10) and (12), neglecting terms of order $o(\|\mathbf{x}-\mathbf{x}_0\|)$, the first-order optimality condition becomes

$$2\mathbf{J}(\mathbf{x}_0)^T \mathbf{r}(\mathbf{x}) \mathbf{J}(\mathbf{x}_0) = 2\mathbf{J}(\mathbf{x}_0)^T (\mathbf{J}(\mathbf{x}_0)\mathbf{x} + \mathbf{a}(\mathbf{x}_0)(\boldsymbol{\delta b} - \mathbf{x}_0) - ). \tag{13}$$

Therefore, if $\mathbf{x}_0$ satisfies the first-order optimality conditions, introducing the perturbation $\boldsymbol{\delta b}$ alters the critical point, and the change can be determined by solving the normal equations

$$\mathbf{J}(\mathbf{x}_0)^T \mathbf{J}(\mathbf{x}_0)(\mathbf{x}-\mathbf{x}_0) = \mathbf{J}(\mathbf{x}_0)^T \boldsymbol{\delta b}, \tag{14}$$

which involve the symmetric matrix $\mathbf{J}(\mathbf{x}_0)^T \mathbf{J}(\mathbf{x}_0)$. This problem is known to have condition number $\kappa$, which depends on the ratio of the largest to the smallest singular value of $\mathbf{J}(\mathbf{x}_0)$, so that this quantity also characterizes the condition number of our nonlinear problem (after first-order truncation). The interested reader is referred to (Demmel, 1997) for further details on the conditioning of linear least square problems. In our approach, this condition number is computed at the minimum point, and the singular values and vectors are also reported (see Tables 1 and 4 for the Clock Tower and Table 7 for the Maddalena bridge).

When the Jacobian at the end of the optimization process is ill-conditioned, the solution cannot be guaranteed to high accuracy. In fact, small perturbation on the input (such as noise) create large perturbations in the retrieved solution. However, a closer look at the structure of these perturbations reveals that some information can still be extracted by the process. With the aim of briefly presenting the theoretical analysis, let us assume, for simplicity, that the Jacobian has only one singular value close to zero, that is, $\sigma_p \approx 0$, but $\sigma_{p-1} >> 0$. The general case does not cause any more difficulties, but this choice makes the notation lighter. Let $\mathbf{q}^{(p)}$ and $\mathbf{z}^{(p)}$ be the corresponding left and right singular vectors. Then, given any uncertainty $\boldsymbol{\delta b}$ on the measured data $\mathbf{b}$, in view of (14), we can write the solution to the perturbed problem as

$$\mathbf{x} = \mathbf{x}_0 + (\mathbf{J}(\mathbf{x}_0)^T \mathbf{J}(\mathbf{x}_0))^{-1} \mathbf{J}(\mathbf{x}_0)^T \, \boldsymbol{\delta b} = \mathbf{x}_0 + \sum_{i=1}^{p} \mathbf{z}^{(i)} \sigma_i^{-1} \mathbf{q}^{(i)T} \boldsymbol{\delta b} \qquad (15)$$

where the last equation follows by replacing $\mathbf{J}(\mathbf{x}_0)$ with its SVD. The important information in equation (15) is that the large perturbations to $\mathbf{x}_0$, which correspond to the small singular values, are all aligned in the directions of the corresponding $\mathbf{z}^{(i)}$. Therefore, even though the computed solution $\mathbf{x}$ cannot be trusted completely, its component along the directions corresponding to large singular values (say, larger than the noise level) can be safely taken into consideration.

For example, let us consider a problem in which $\mathbf{x}$ belongs to a two-dimensional space, and assume we compute a solution whose last singular value is close to zero, whereas the other is non-negligible. If we notice that $\mathbf{z}^{(2)}$ is parallel to $\mathbf{x}$, then this means that for moderate values of $\beta$, $(1+\beta)\mathbf{x}$ is a good solution as well. Nevertheless, the ratio between the two parameters $x_1/x_2$ can be trusted to be computed accurately, and will not be affected by small perturbations in the data.

This situation is often encountered when modeling a homogeneous material and optimizing over its Young's modulus and mass density. Since the frequencies and the eigenmodes are invariant for rescaling of these two parameters, the optimization process can reliably determine their ratio, but not their exact value. Later on, we will demonstrate this simple consideration through a practical example (the Clock Tower). However, it should be emphasized that the same approach can uncover unexpected relations, which in more complicated cases may involve several parameters.

In practice, when the level of noise is known a priori, (15) shows that we can trust the components of the solution corresponding to singular values larger than the noise level.

**Case studies**

In this section we verify the performance of our approach in three example applications. First, a very simple numerical application, an arch on piers, is used to assess the correctness and robustness of the optimization algorithm. Then, the proposed method is applied to two age-old masonry structures, whose response to ambient vibrations has been measured experimentally: the Clock Tower in Lucca and the Maddalena Bridge in Borgo a Mozzano, described and analyzed in Pellegrini et al. 2017, Azzara et al. 2017 and Girardi et al. 2018 are revisited herein. In particular, the experimental results obtained by applying system identification techniques (Stochastic Subspace Identification and Enhanced Frequency Domain Decomposition methods, Reynders et al., 2014) to the velocities recorded during monitoring campaigns conducted in 2015-2016 are used to solve the minimum problem (5), with objective function (6) and then to calibrate the FE models.

For the sake of simplicity, in these examples only few parameters are allowed to vary: no substantial changes in the proposed algorithm are necessary to take into account more general cases, including unknown boundary conditions, as shown in Girardi et al., 2018b, De Falco et al. 2018.

The analyses presented in this section have been performed on a computer with an Intel Core i7-920 CPU running at 2.67 GHz and 18 GB of RAM clocked at 1066 MHz; the convergence tolerance was set to $10^{-3}$, that is, the trust-region procedure is stopped when the norm of the projected gradient falls below $10^{-3}$ (Girardi et al. 2018 a).

*The arch on piers*

Let us discuss a first, simple example for demonstration purposes. Consider the plane model of a masonry arch on piers, whose FE discretization is shown in Figure 2. The arch spans 4 m, and rests on two 4m - high lateral piers, clamped at the base. The structure is modeled by means of 336 finite four-node plane strain elements (element 6 in Binante et al., 2017), for a total number of 851 degrees of freedom. Three different materials, depicted with different colors in Figure 2, compose the structure: the arch is made up of material 1 (green in the figure), while materials 2 and 3 compose the piers (red and blue, respectively). The Poisson's ratio is set to 0.2 for the whole structure and remains constant during the numerical experiments.

We first run the NOSA-ITACA code by assigning the mechanical properties to all the constituent materials and calculate the model's eigenfrequencies. For the three materials, let us assume the following values of the Young's modulus and mass densities:

$$E_1 = 3.25 \text{ GPa}, \quad E_2 = 5.00 \text{ GPa}, \quad E_3 = 4.80 \text{ GPa} \tag{16}$$

$$\rho_1 = 1800 \frac{\text{kg}}{\text{m}^3}, \quad \rho_2 = 2200 \frac{\text{kg}}{\text{m}^3}, \quad \rho_3 = 2100 \frac{\text{kg}}{\text{m}^3} \tag{17}$$

where the index $j \in \{1,2,3\}$ indicates the material under consideration. The first 5 eigenfrequencies of the model evaluated by the NOSA-ITACA code are:

$$\mathbf{f} = [9.575, 14.87, 23.17, 39.17, 62.84] \text{ Hz} \tag{18}$$

Then, in order to test the algorithm, we assume that the values of $E_2$, $E_3$, and $\rho_3$ are unknown and varying within the intervals

$$1\text{GPa} \leq E_2 \leq 9\text{GPa}, \quad 1\text{GPa} \leq E_3 \leq 9\text{GPa}, \quad 1000 \frac{\text{kg}}{\text{m}^3} \leq \rho_2 \leq 3000 \frac{\text{kg}}{\text{m}^3}, \tag{19}$$

which define the box $\Omega$ in (3). Given the frequency vector (18), we aim at finding, through the optimization algorithm, the optimal values of $E_2$, $E_3$, and $\rho_3$ and comparing them with (16), (17). In the default implementation, the starting points are the midpoints of the intervals. Convergence in this case is shown in Figure 3, where each iteration corresponds to building a reduced model. The starting points are sufficiently close to the correct ones, and the solution is very close to the optimum, as shown in the right graph. The left plot in Figure 3, which shows the convergence history of the objective function in log-scale, demonstrates that we are recovering the right solution. To test the robustness of the approach, we run again the algorithm with new starting points

$$E_2^{(0)} = 2 \text{ GPa}, \ E_3^{(0)} = 1.10 \text{ GPa}, \ \rho_2^{(0)} = 1100 \frac{\text{kg}}{\text{m}^3} \tag{20}$$

Convergence of the method is now displayed in Figure 4: the initial frequencies are quite far from the correct ones, nevertheless the method still reaches convergence by iterating on 11 reduced models; moreover, a good estimate is already obtained in 6 steps. The optimal values of $E_2$, $E_3$, and $\rho_3$ calculated for both the initial points coincide with the values in (16), (17).

To test the robustness of the algorithm when the input is affected by noise, as is the case of natural frequencies and mode shapes obtained through experimental data, we perturb the frequency vector $\mathbf{f}^p = \mathbf{f} + \delta\mathbf{f}$ by imposing $\delta f_i \leq |f_i| \cdot \delta$, with $\delta$ the prescribed noise level. Figure 5 plots the results of the numerical tests performed by varying the noise level within the interval $0.01\% \leq \delta \leq 100\%$. The relative error in the figure is defined as the infinity norm of the vector with components $(x_i^p - x_i)/x_i^p$, where $x_i^p$ is the vector of the actual model parameters, evaluated with perturbed values of the frequencies, and $x_i$ is the one corresponding to $\mathbf{f}$. The figure clearly shows that the error is bounded and linearly increases with the noise level.

*The Clock Tower*

The skyline of the ancient city of Lucca is characterized by a number of masonry towers and their belfries, many dating back to the medieval period. Among these monuments, the Clock Tower (Torre delle Ore) (Figure 6) is one of the best known and most visited, thanks to the peculiar shape of its bell chamber, which is clearly visible and recognizable throughout the entire historic centre. Built in the 13$^{th}$ century by local families, since the last decade of the 14$^{th}$ century the Clock Tower has been used as a civic building, taking its name from the big clock visible on its southern facade. The Clock Tower rises at the corner between the roads named Via Arancio and Via Fillungo, one of the most popular in Lucca's historic centre; the adjacent buildings abut the tower on two sides for a height of about 13 m and constitute asymmetric boundary conditions. The Clock Tower is 48.4 m high; it has a rectangular cross section of about 5.1 x 7.1m and walls of thickness varying from about 1.77 m at the base to 0.85 m at the top. Two barrel vaults are set inside the tower at heights of about 12.5 and 42.3 m. The bell chamber, made up of four masonry pillars connected by elliptical arches, stands on the upper barrel vault and is covered by a pavilion roof constructed of wooden trusses and rafters. With regard to the materials constituting the masonry tower, visual inspection reveals that the masonry from the base up to a height of 15 m is made up of regular stone blocks and thin mortar joints. The upper walls are instead made up of regular stone blocks and bricks, also with thin joints. The pillars of the bell chamber are made up of bricks.

On 25 November, 2016 the ambient vibrations of the Clock Tower were monitored for a few hours via four SARA SS20 three-axial seismometric stations made available by the Seismological Observatory of Arezzo (INGV). The instruments were moved along the tower's height by adopting three different layouts, and combining data in order to identify four natural vibration frequencies and mode shapes of the tower (Pellegrini et al, 2017).

Two FE models of the Clock Tower (Figure 7) are used here to test the algorithm proposed in Girardi et al. 2018a and recalled in the previous section; the former (mesh 1) is made of only one material, and the latter (mesh 2) by two different materials indicated in red and green. Each model employs 11383 eight-node brick elements for the masonry (element 8 in Binante et al. 2017), while the tie rods and the wooden roof are modeled via beam elements (element 9 in Binante et al. 2017), for a total number of 45511 degrees of freedom. The mechanical characteristics chosen for the iron and wood components are

$$E_i = 210 \text{ GPa}, \ \rho_i = 7850 \frac{\text{kg}}{\text{m}^3}, \ \nu_i = 0.3, \ E_w = 8 \text{ GPa}, \ \rho_w = 800 \frac{\text{kg}}{\text{m}^3}, \ \nu_w = 0.35. \tag{21}$$

The minimum of function (6) is calculated for the two models by choosing $w_i = \hat{f}_i^{-1}$, $i = 1...4$, $w_5 = w_6 = 0.1$, $w_7 = w_8 = 0$.

Let us begin by considering the Clock Tower as composed of one single material (mesh 1) with Poisson's ratio fixed at 0.2, while Young's modulus $E_1$ and mass density $\rho_1$ vary in the ranges

$$1.5 \text{ GPa} \leq E_1 \leq 3.5 \text{ GPa}, \ 1000 \ \frac{\text{kg}}{\text{m}^3} \leq \rho_1 \leq 2500 \frac{\text{kg}}{\text{m}^3}. \tag{22}$$

The plot of function $\Phi(E_1, \rho_1)$, whose minimum as calculated by our numerical procedure is attained at $E_{1opt} = 2.589$ GPa, $\rho_{1opt} = 1700 \frac{\text{kg}}{\text{m}^3}$, is reported in Figure 8. The choice of the ends of the intervals in (22) is driven by engineering judgment. We stress that, whenever the final optimum points are enclosed in the internal of the intervals (as happens in all the considered case studies), the role of the box constraints is not strong: the same solution would have been found allowing the parameters to move on the entire positive real line, assuming the use of the same starting point.

To ease interpretation of the results, here and in the following we consider the Jacobian of the function $\Phi$ whose input parameters have been scaled so that the minimum is attained when all of them are equal to 1. The SVD of the scaled Jacobian $\mathbf{J}(E_{1opt}, \rho_{1opt})$ of the residual function defined in (8) yields the results summarized in Table 1, with the singular values $\sigma_1 > \sigma_2$ reported in the first column and the corresponding right singular vectors in the second and third columns. The values in Table 1 are in agreement with Figure 8, where $\Phi(E_1, \rho_1)$ exhibits a "flat" direction parallel to $\mathbf{z}^{(2)}$, corresponding to the singular value $\sigma_2 = 0.0054255$.

This is an instance of the situation described in the previous section, in which the solution to the optimization problem is far from being unique, and the numerical procedure is only able to calculate the optimal ratio between Young's modulus and mass density. Indeed, note that the non-negligible singular value has a singular vector almost parallel to the vector $\begin{bmatrix} -1 & 1 \end{bmatrix}^T$, and the approximate null-space is generated by the vector with components $\begin{bmatrix} 1 & 1 \end{bmatrix}^T$. This implies that increasing the parameters while keeping their ratio constant causes very little change in the objective function. The relatively large condition number $\kappa = \dfrac{\sigma_1}{\sigma_2} = 202.9$ reveals the high dependence of the parameters on the noise that may affect the experimental data.

The availability of information on the material constituting the structure can help overcome the problems connected with the non-uniqueness of the solution, and allows the number of unknown parameters to be reduced. Thus, if we fix the mass density to $\rho = 2100 \text{ kg/m}^3$, a reasonable value for the material constituting the tower, and update the Young's modulus

$$1 \text{ GPa} \leq E \leq 6 \text{ GPa}, \tag{23}$$

the (unique) minimum point of the objective function $\Phi(E)$, plotted in Figure 9, is then $E_{opt} = 3.183 \text{ GPa}$. Figure 10 shows the convergence of the model's frequencies to the experimental values during the process, and Table 2 reports the experimental frequencies, the numerical frequencies calculated for $E = E_{opt}$ and the MAC indicators $\gamma_i$. Figure 11 and 12 show the experimental mode shapes (from 1 to 4 from the left) and the numerical mode shapes corresponding to $E = E_{opt}$, calculated via the NOSA-ITACA code.

As far as mesh 2 is concerned, two materials are involved: material 1 for the bell chamber and material 2 for the tower (respectively red and green in Figure 7). Poisson's ratio is 0.2 and the mass densities are

$$\rho_1 = 1700 \text{ kg/m}^3, \quad \rho_2 = 2100 \text{ kg/m}^3. \tag{24}$$

We update Young's moduli $E_1$ and $E_2$, which vary in the ranges

$$1 \text{ GPa} \leq E_1 \leq 6 \text{ GPa}, \qquad 1 \text{ GPa} \leq E_2 \leq 6 \text{ GPa}. \qquad (25)$$

In this case the minimum point of $\Phi$ has coordinates $E_{1opt} = 1.9288$ GPa, $E_{2opt} = 3.0451$ GPa, which correspond to the Young's modulus of full brick and lime mortar, and stone masonry, respectively, according to the Italian technical guidelines (M.I.T. 2009).

Figure 13 shows a plot of the objective function $\Phi(E_1, E_2)$, while Figure 14 shows the convergence of the model's frequencies to the experimental values during the process. Table 3 reports the experimental frequencies, the numerical frequencies calculated at the minim point $(E_{1opt}, E_{2opt})$ and the MAC indicators $\gamma_i$ as well.

The SVD of the Jacobian $\mathbf{J}(E_{1opt}, E_{2opt})$ of the residual function defined in (8) yields the results summarized in Table 4, which are in agreement with Figure 13, where $\Phi(E_1, E_2)$ exhibits a unique minimum point. From Table 4, we get the condition number $\kappa = 3.7$, which, unlike that in Table 1, is quite low.

A comparison between the results reported in Tables 2 and 3 shows that, with respect to the one material mesh, using two materials yields lower relative errors on the frequencies. Moreover, the third and fourth numerical frequencies, which are very close in Figure 10 for mesh 1, tend to separate and approach their experimental counterparts in Figure 14 for the mesh 2. Lastly, with regard to mode shapes, the MAC indicator $\gamma_4$ increases from 0.6041 for mesh 1, to 0.8950 for mesh 2, showing that the non homogeneous model is more realistic. The parallelism of the experimental mode shapes $\hat{\mathbf{v}}_3$ and $\hat{\mathbf{v}}_4$ (their MAC indicator is 0.9072) suggests that measurement of $\hat{\mathbf{v}}_3$ is not sufficiently accurate, and could explain the low value of $\gamma_3$ in Table 3.

For this case, the software had already been used in Girardi et al. 2018a, where only the frequencies were optimized (the eigenmode matching was omitted), and the parameters and frequencies identified coincide nonetheless. In fact, good conditioning of this problem (even ignoring the eigenmodes, the condition number of the Jacobian does not exceed 4) enables the unknown parameters to be determined effectively using the frequencies alone.

The total computation time was 22.54 s for mesh 1 and 27.15 s for mesh 2. Instead, the time required to solve the minimum problem (5)-(6) with a general purpose optimizer (using NOSA-ITACA as a black box function and the SQP solver from the Optimization Toolbox of Matlab) is 39.55 s for mesh 1 and 175.41 s for mesh 2. In the latter case we made use of precomputed parametric assembly, and therefore the timings only include computation of the eigenvalues and eigenvectors at any point required by the optimizer. Table 5 compares the performance of the two algorithms in terms of optimal values of the parameters, frequencies and computation times for mesh 2. For mesh 1 the results obtained via the two algorithms coincide and are omitted.

*The Maddalena bridge*

Also known as the Devil's Bridge, this fascinating structure (see Fig. 15) can be dated back to around the 11th century and crosses the Serchio River for a total length of about 100 meters, deviating over fifteen degrees from the direction perpendicular to the river. It consists of one large arch of 38 m in span and three smaller arches on the left-hand side. The main arch, which is just one meter high at the key, has a perfectly circular intrados profile and springs from the rock of the riverbed. The transverse section of the bridge ranges between 3.5 m and 3.7 m.

A detailed historical review of the bridge has been carried out by Gucci and De Falco 2010, and a finite element model aimed at studying its static and dynamic behavior has been proposed in De Falco et al. 2014 and is shown in Figure 16. The model consists of 41954 eight-node brick elements, for a total of 155312 degrees of freedom. In June 2015 an experimental campaign was conducted to measure the ambient vibrations acting on the bridge. To this end, four SARA SS20 seismometric stations were used, arranged in different layouts over the bridge during five different tests. Details on these measurements are given in Azzara et al., 2017. Table 6 reports the values of the bridge's first six natural frequencies obtained from the experimental data and averaged over the five tests. The corresponding experimental mode shapes were extracted from the data and are shown in Figure 17 (Azzara et al., 2017).

With regard to the constituent materials, the external surface of the bridge is for the most part made of a particular variety of sandstone (Macigno sandstone) and, to a lesser extent, of a blue limestone, with tight uniform joints. Thanks to the results of a seismic refraction survey (Gucci and De Falco, 2010), some information is also available on the mechanical characteristics of the materials under the bridge's stone pavement. No direct information is however available on the exact composition of the bridge's cross section, the thickness of the stone layers, or the geometry and mechanical characteristics of its inner portions. In this paper, two different hypotheses have been considered regarding the materials distribution. The first, mesh 1 (Figure 16), considers a unique, homogeneous material, with unknown properties, while the second, mesh 2 (Figure 18), accounts for two different materials – the external stone layer (assumed to be about 0.5 m thick) and an infill with unknown properties. For both models the minimum of the objective function (6) is calculated choosing $w_i = \hat{f}_i^{-1}, i = 1...6$, $w_7 = 0.1$, $w_j = 0, j = 8...12$, that is, considering only the bridge's first mode shape, in whose evaluation we are quite confident.

Let us begin with mesh 1. In order to avoid the situation encountered in the Clock Tower example, with a nearly infinite number of possible solutions, we fix the mass density of the material at $\rho = 1900 \frac{\text{kg}}{\text{m}^3}$. We also set Poisson's ratio at 0.2 and the stiffness of the springs modeling the soil under the central pier of the bridge at $k_s = 1.929 \cdot 10^{10} \text{ N/m}^3$. This value comes from a preceding analysis conducted in De Falco et al. 2018, in which the optimal value of the springs' stiffness has been calculated via the proposed algorithm. The bridge's Young's modulus is then allowed to vary within the interval

$$4 \text{ GPa} \leq E \leq 15 \text{ GPa}. \tag{26}$$

The plot of function $\Phi(E)$, whose minimum is attained at $E_{opt} = 7.3951\,\text{GPa}$, is reported in Figure 19. As pointed out in Azzara et al. 2017, the high resulting value of the Young's modulus can be regarded as an initial, dynamic value of the high quality masonry constituting the Maddalena bridge. Figure 20 shows the numerical mode shapes (from 1 to 6 from the left) of the bridge model corresponding to $E = E_{opt}$, calculated via the NOSA-ITACA code. Table 6 reports the experimental frequencies, the numerical frequencies calculated for $E = E_{opt}$ and the MAC indicators $\gamma_i$. The table highlights the error in the estimation of the fundamental frequency (on the order of 6%): the bridge turns out to be stiffener than the model for out-of-plane movements. With regard to the estimation of the bridge's mode shapes, the MAC indicators show that the first and fifth mode shapes are matched with high confidence. Instead, with regard to the fourth mode, the numerical and experimental eigenvectors seem to be orthogonal.

As far as mesh 2 is concerned, the three parameters to be updated are the Young's modulus $E_1$ of the bridge's external stone surface, the Young's modulus $E_2$ and mass density $\rho_2$ of the bridge's infill material. The mass density of the stone surface is fixed at $\rho_1 = 2000\,\text{kg/m}^3$, Poisson's ratio is 0.2; the buttresses strengthening the bridge piers in the riverbed have fixed material properties, $\rho_3 = 1900\,\text{kg/m}^3$ and $E_3 = 7\,\text{GPa}$. We allow the parameters to vary within the intervals

$$4\,\text{GPa} \leq E_1 \leq 15\,\text{GPa}, \quad 1200\,\frac{\text{kg}}{\text{m}^3} \leq \rho_2 \leq 2000\,\frac{\text{kg}}{\text{m}^3}, \quad 1\,\text{GPa} \leq E_2 \leq 15\,\text{GPa} \tag{27}$$

In this case, the minimum point of $\Phi$ has coordinates

$$E_{1opt} = 10.381\,\text{GPa}, \quad E_{2opt} = 4.714\,\text{GPa}, \quad \rho_{2opt} = 1960\,\frac{\text{kg}}{\text{m}^3}. \tag{28}$$

With regard to the mass density $\rho_{2opt}$, the value yielded by the algorithm is very close to the fixed value of the external surface. The value $E_{1opt}$ is acceptable for sandstone and limestone masonry with very thin mortar joints (the compressive strength of the Macigno sandstone constituting the main part of the bridge's visible masonry ranges from 0.1 GPa to 0.14 GPa). Modulus $E_{2opt}$ reflects the results of the sonic tests, yielding an average velocity of about 1500 m/s within the infill masonry (the corresponding value of Young's modulus is about 4.2 GPa for a density of 2000 kg/m$^3$) (Azzara et al., 2017). Therefore, the optimal values found are in good agreement with the information available on the structure. Nevertheless, the errors in the estimation of the bridge's frequencies (Table 7) are essentially the same as those for the equivalent model of mesh 1 (Table 6). This behavior could be explained by Figure 21, which shows a plot of the objective function $\Phi(E_1, E_2, \rho_2)$ for $\rho_2 = \rho_{2opt}$. The function clearly exhibits a "flat" direction and, if moduli $E_1$ and $E_2$ are changed along that direction, the solution, in terms of the bridge's estimated modal properties, is almost equally satisfactory. This is also confirmed by Table 8, which shows the results of the singular value decomposition of $\mathbf{J}(E_{1opt}, E_{2opt}, \rho_{2opt})$: the "flat" direction turns out to be parallel to $\mathbf{z}^{(3)}$, which corresponds to the smallest singular value $\sigma_3 = 0.012566$ (the condition number in this case is $\kappa = 68.29$). In this case, the right singular vector $\mathbf{z}^{(3)}$ represents the direction along which the global stiffness of the bridge's section remains essentially unchanged, when $E_1$ and $E_2$ vary. Note that $\mathbf{z}^{(3)}$ entails a diminishing modulus $E_1$ as modulus $E_2$ increases (as also depicted in Figure 21): therefore, the homogeneous case, with $E_1 = E_2$, is also included.

To conclude, mesh 2 provides more realistic values of the bridge's constituent materials. However, for both models, the out-of-plane stiffness of the model is underestimated. The reason for this finding, as highlighted in (Manos et al., 2016) for similar bridge types in Greece, may lie in the anisotropic behavior of the bridge's masonry, which is not taken into account in the model.

With regard to the mode shapes, Tables 6 and 7 highlight the low values of the MAC indicators, with the exception of the first and fifth mode shapes. One explanation for this may be the low accuracy of the experimental mode shapes, which were acquired in only a few locations along the bridge. Finally, Figure 22 shows the convergence history of the frequencies calculated for mesh 1 (on the left) and mesh 2.

The total computation time was 182.96 s for mesh 1 and 398.11 s for mesh 2. The time required to solve the minimum problem (5)-(6) with a general purpose optimizer, using NOSA-ITACA as a black box function and the SQP solver (as for the Clock Tower, for which precomputed parametric assembly was exploited), was 471.61 s for mesh 1 and 2115.8 s for mesh 2. Tables 9 and 10 compare the performance of the two algorithms in terms of optimal values of the parameters, frequencies and computation times for mesh 1 and mesh 2, respectively.

The experimental data are provided in Azzara et al. 2017, along with their variance. The relative accuracy reported for the measured frequencies is in the range $\left[10^{-3}, 10^{-2}\right]$ and in particular the error magnitude is bounded by the singular values in Table 8. We can use this observation to conclude that the computed parameters are not particularly affected by noise, with the single possible exception of perturbation in the $\mathbf{z}^{(3)}$ direction, whose singular value is the only one close to the noise level.

We shall now present a simple, yet representative example to clarify the analysis. We changed the input data (both the frequencies and the mode shapes under consideration) with a relatively small random perturbation, introducing an error bounded by 1% (in norm). Running the optimizer with these new data yields the optimal parameters:

$$\hat{E}_{1opt} = 10.5 \text{ GPa}, \ \hat{E}_{2opt} = 4.652 \text{ GPa}, \ \hat{\rho}_{2opt} = 1958 \frac{\text{kg}}{\text{m}^3}.$$

The vector containing the relative perturbations of the parameters is given by $\mathbf{z} \approx \begin{bmatrix} -0.0115 & 0.0133 & 0.0014 \end{bmatrix}$ and $\|\mathbf{z}\| \approx 1.75 \cdot 10^{-2}$. Therefore, the noise has caused less change than the worst-case bound, given by the ratio of the largest to the smallest singular values of the Jacobian (about 68 in this case). However, decomposing vector $\mathbf{z}$ into the directions $\mathbf{z}^{(1)}$, $\mathbf{z}^{(2)}$ and $\mathbf{z}^{(3)}$ yields

$$\mathbf{z} = \zeta_1 \mathbf{z}^{(1)} + \zeta_2 \mathbf{z}^{(2)} + \zeta_3 \mathbf{z}^{(3)}, \quad \zeta = \begin{bmatrix} 0.0042 \\ 0.0027 \\ 0.0168 \end{bmatrix}.$$

Note that the largest component of the perturbation is in the direction of $\mathbf{z}^{(3)}$, which in fact corresponds to the smallest singular value. This has the consequence that the relative error on the mass density is smaller, according to the size of the third component of $\mathbf{z}^{(3)}$. Once more, this simple example shows how the component of the singular vectors can be used to assess the expected accuracy and robustness of the retrieved parameters.

**Concluding remarks**

In this work, we have investigated the use of model updating techniques to determine unknown parameters in historical masonry buildings.

We propose a model updating algorithm, which relies on the use of local parametric reduced-order models embedded in a trust region scheme. The objective function to be minimized measures the discrepancy between the computed and experimental data, in terms of both natural frequencies and mode shapes. The algorithm, implemented in the NOSA-ITACA framework, turns out to be very efficient with respect to commercial optimizers and, unlike black-box tools, allows evaluating the reliability of the solution.

In the paper, a simple preliminary application is presented to test the correctness and robustness of the numerical method. Then, the algorithm is applied on two historical buildings: the Clock Tower in Lucca and the Maddalena Bridge in Borgo a Mozzano, whose experimental dynamic properties were at our disposal. Particular attention has been devoted to the choice of the parameters, by considering several scenarios for the structures' constituent materials, and we have discussed the conditioning and well-posedness of the problem.

The singular value decomposition of the Jacobian of the objective function at the minimum point allows interpreting the results and assessing their reliability and sensitivity to noisy experimental data. A natural line of research stemming from this analysis is the use of regularization techniques to filter the noise, thereby automating the retrieval of meaningful solutions.

The influence of eigenmodes on the minimization process seems not to be relevant in the considered case studies, in which mode shapes have been measured with lower accuracy than the frequencies, and further investigations on this issue are planned.

*Acknowledgements*

This research has been partially supported by the Region of Tuscany and MIUR, the Italian Ministry of Education, Universities and Research, within the Call FAR-FAS 2014 (MOSCARDO Project: ICT technologies for structural monitoring of age-old constructions based on wireless sensor networks and drones, 2016-2018) and by the GNCS/INdAM project "Metodi numerici avanzati per equazioni e funzioni di matrici con struttura". These supports are gratefully acknowledged.

Table 1. Clock Tower, mesh 1: singular values and right singular vectors of the scaled Jacobian $\mathbf{J}(E_{1opt}, \rho_{1opt})$

| $\sigma$ | $\mathbf{z}^{(1)}$ | $\mathbf{z}^{(2)}$ |
|---|---|---|
| 1.1009 | -0.71635 | 0.69774 |
| 0.0054255 | 0.69774 | 0.71635 |

Table 2. Clock Tower, mesh 1, results of the optimization algorithm: frequencies, relative errors with respect to the experimental values and MAC indicators

|  | Experimental frequencies [Hz] | Numerical frequencies [Hz] | Relative errors [%] | $\gamma_i$ |
|---|---|---|---|---|
| **Mode shape 1** | **1.05** | 1.0073 | 4.00 | 0.986 |
| **Mode shape 2** | **1.3** | 1.2683 | 2.44 | 0.9779 |
| **Mode shape 3** | **4.19** | 4.4337 | 5.82 | 0.5982 |
| **Mode shape 4** | **4.50** | 4.4917 | 0.18 | 0.6041 |

Table 3. Clock Tower, mesh 2, results of the optimization algorithm: frequencies, relative errors with respect to the experimental values and MAC indicators

|  | Experimental frequencies [Hz] | Numerical frequencies [Hz] | Relative errors [%] | $\gamma_i$ |
| --- | --- | --- | --- | --- |
| **Mode shape 1** | **1.05** | 1.0440 | 0.57 | 0.9862 |
| **Mode shape 2** | **1.3** | 1.3129 | 0.99 | 0.9782 |
| **Mode shape 3** | **4.19** | 4.1884 | 0.04 | 0.2731 |
| **Mode shape 4** | **4.50** | 4.4522 | 1.06 | 0.8950 |

Table 4. Clock Tower, mesh 2: Singular values and right singular vectors of the scaled Jacobian $\mathbf{J}(E_{1opt}, E_{2opt})$

| $\sigma$ | $\mathbf{z}^{(1)}$ | $\mathbf{z}^{(2)}$ |
|---|---|---|
| 0.67014 | -0.17148 | 0.98519 |
| 0.18113 | -0.98519 | -0.17148 |

Table 5. Clock Tower, Mesh 2. Comparison between the proposed algorithm (reduced model, RM) and a general purpose optimizer (GPO): optimal parameters, frequencies and computation times.

| Numerical frequencies [Hz] | | Optimal parameters [GPa] | |
|---|---|---|---|
| **RM** | **GPO** | **RM** | **GPO** |
| 1.0440 | 1.0441 | $E_1 = 1.9288$ | $E_1 = 1.9309$ |
| 1.3129 | 1.3130 | $E_2 = 3.0451$ | $E_2 = 3.0444$ |
| 4.1884 | 4.2067 | **Computation times [s]** | |
| 4.4522 | 4.4611 | 27.15 | 175.41 |

Table 6. Maddalena Bridge, mesh 1, results of the optimization algorithm: frequencies, relative errors with respect to the experimental values and MAC indicators

|  | Experimental frequencies [Hz] | Numerical frequencies [Hz] | Relative errors [%] | $\gamma_i$ |
|---|---|---|---|---|
| **Mode shape 1** | **3.37** | 3.1696 | 5.95 | 0.988 |
| **Mode shape 2** | **5.06** | 5.2409 | 3.58 | 0.298 |
| **Mode shape 3** | **5.40** | 5.508 | 2.00 | 0.501 |
| **Mode shape 4** | **7.06** | 6.8738 | 2.64 | 0.078 |
| **Mode shape 5** | **8.80** | 8.6522 | 1.68 | 0.973 |
| **Mode shape 6** | **9.19** | 9.7321 | 5.90 | 0.485 |

Table 7. Maddalena Bridge, mesh 2, results of the optimization algorithm: frequencies, relative errors with respect to the experimental values and MAC indicators

|  | Experimental frequencies [Hz] | Numerical frequencies [Hz] | Relative errors [%] | $\gamma_i$ |
|---|---|---|---|---|
| **Mode shape 1** | **3.37** | 3.1863 | 5.45 | 0.9873 |
| **Mode shape 2** | **5.06** | 5.2334 | 3.43 | 0.3037 |
| **Mode shape 3** | **5.40** | 5.4268 | 0.50 | 0.5063 |
| **Mode shape 4** | **7.06** | 7.1665 | 2.36 | 0.0497 |
| **Mode shape 5** | **8.80** | 8.6327 | 1.90 | 0.9727 |
| **Mode shape 6** | **9.19** | 9.7318 | 5.90 | 0.4756 |

Table 8. Maddalena bridge, mesh 2: singular values and right singular vectors of the scaled Jacobian $\mathbf{J}(E_{1opt}, E_{2opt}, \rho_{2opt})$

| $\sigma$ | $\mathbf{z}^{(1)}$ | $\mathbf{z}^{(2)}$ | $\mathbf{z}^{(3)}$ |
|---|---|---|---|
| 0.85818 | -0.62465 | -0.66249 | -0.41342 |
| 0.16877 | -0.29574 | -0.28929 | 0.91041 |
| 0.012566 | 0.72274 | -0.69095 | 0.015217 |

Table 9. Maddalena bridge, mesh 1. Comparison between the proposed algorithm (reduced model, RM) and a general purpose optimizer (GPO): optimal parameters, frequencies and computation times.

| Numerical frequencies [Hz] | | Optimal parameters [GPa] | |
|---|---|---|---|
| **RM** | **GPO** | **RM** | **GPO** |
| 3.1696 | 3.1708 | $E = 7.3951$ | $E = 7.3946$ |
| 5.2409 | 5.2427 | | |
| 5.508 | 5.5084 | | |
| 6.8738 | 7.3043 | **Computation times [s]** | |
| 8.6522 | 8.6552 | 182.96 | 471.61 |
| 9.7321 | 9.7353 | | |

Table 10. Maddalena bridge, mesh 2. Comparison between the proposed algorithm (reduced model, RM) and a general purpose optimizer (GPO): optimal parameters, frequencies and computation times.

| Numerical frequencies [Hz] | | Optimal parameters [GPa, kg/m$^3$] | |
|---|---|---|---|
| **RM** | **GPO** | **RM** | **GPO** |
| 3.1863 | 3.1886 | $E_1 = 10.5$ | $E_1 = 10.475$ |
| 5.2334 | 5.2356 | $E_2 = 4.652$ | $E_2 = 4.626$ |
| 5.4268 | 5.4242 | $\rho_2 = 1958$ | $\rho_2 = 1960$ |
| 7.1665 | 7.1697 | **Computation times [s]** | |
| 8.6327 | 8.6381 | 398.11 | 2115.8 |
| 9.7318 | 9.7382 | | |

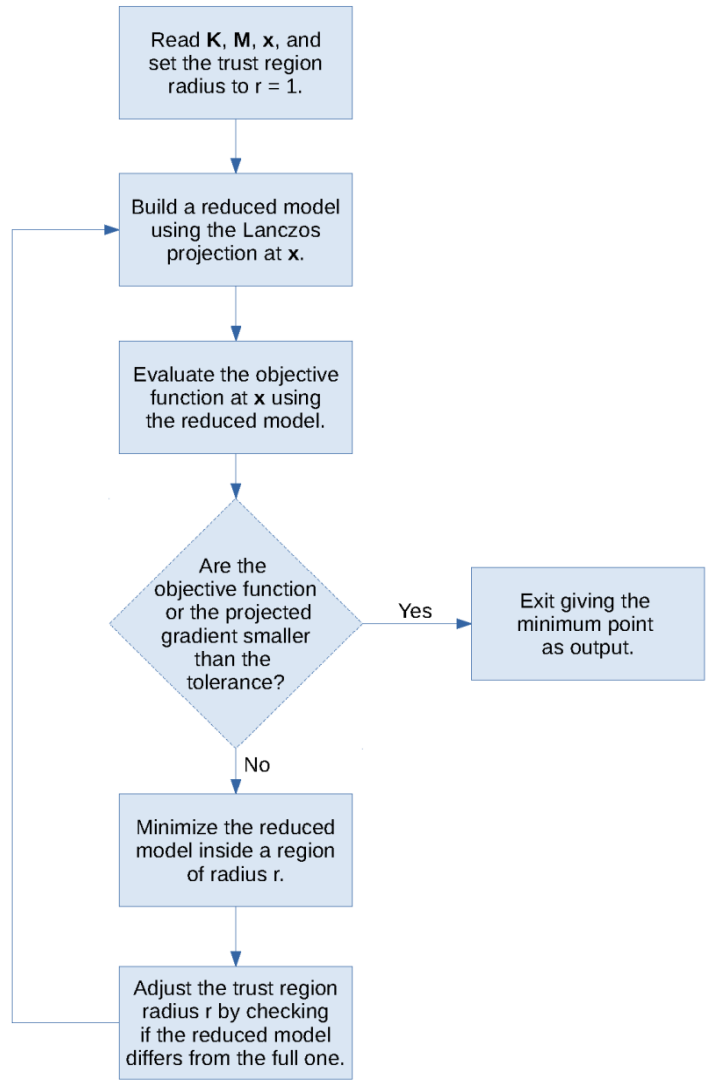

Figure 1. Flowchart of the proposed algorithm.

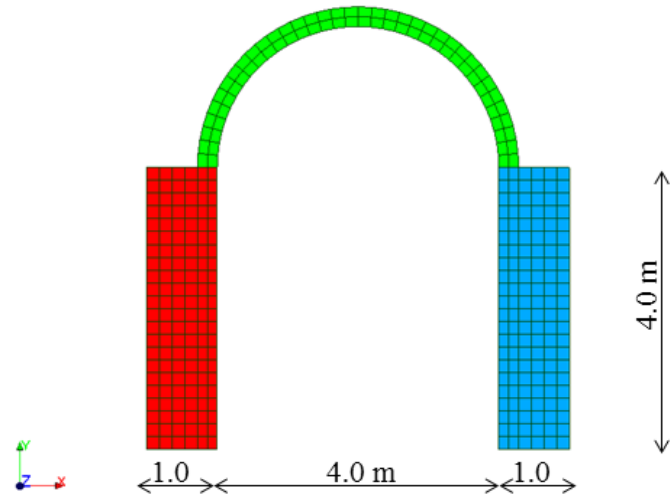

Figure 2. The arch on piers. Each differently colored region corresponds to a different material.

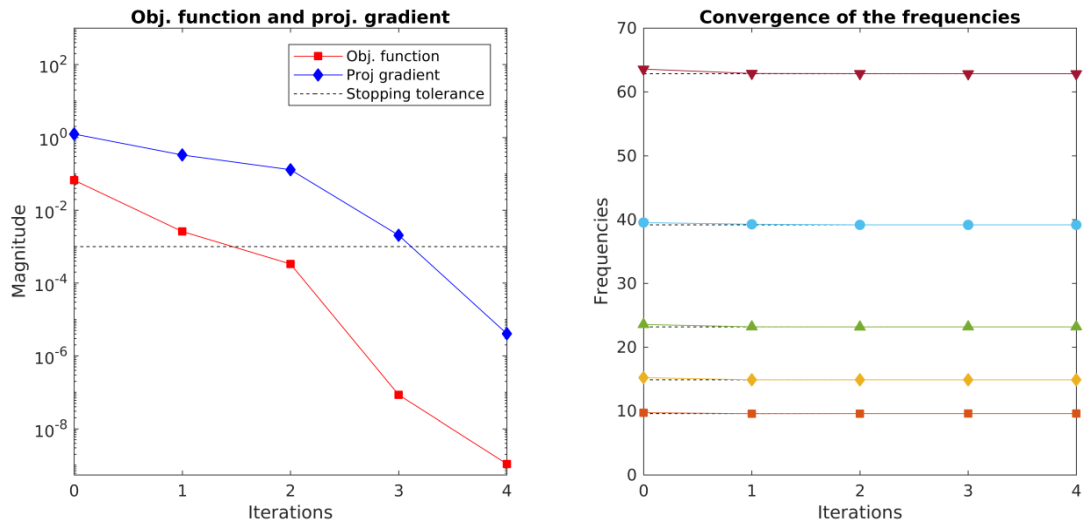

Figure 3. Convergence history of the objective function to the minimum (on the left). The dashed line is the tolerance set for the projected gradient in the optimization scheme. Convergence of the frequencies during the process (on the right) vs. number of reduced models. The starting point is the midpoint of $\Omega$.

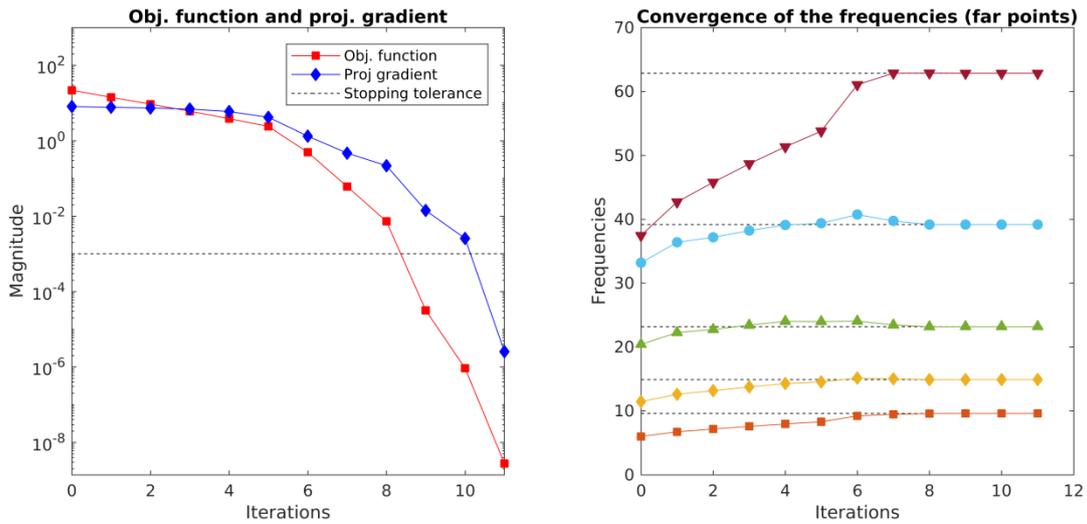

Figure 4. Convergence history of the objective function to the minimum (on the left). The dashed line is the tolerance set for the projected gradient in the optimization scheme. Convergence of the frequencies during the process (on the right) vs. number of reduced models. The starting point is given in (20).

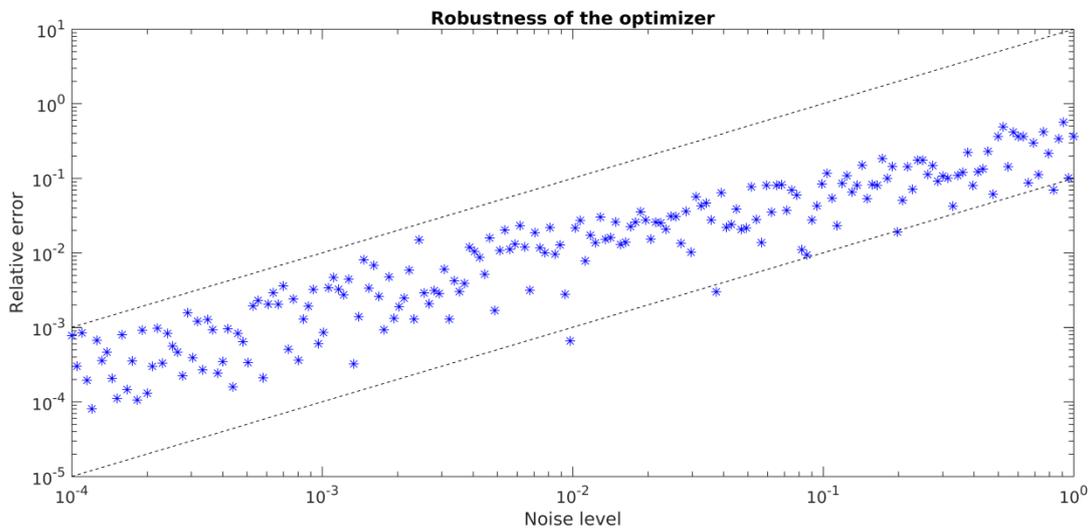

Figure 5. Relation between the noise of the measured frequencies and the optimal parameters calculated by the algorithm.

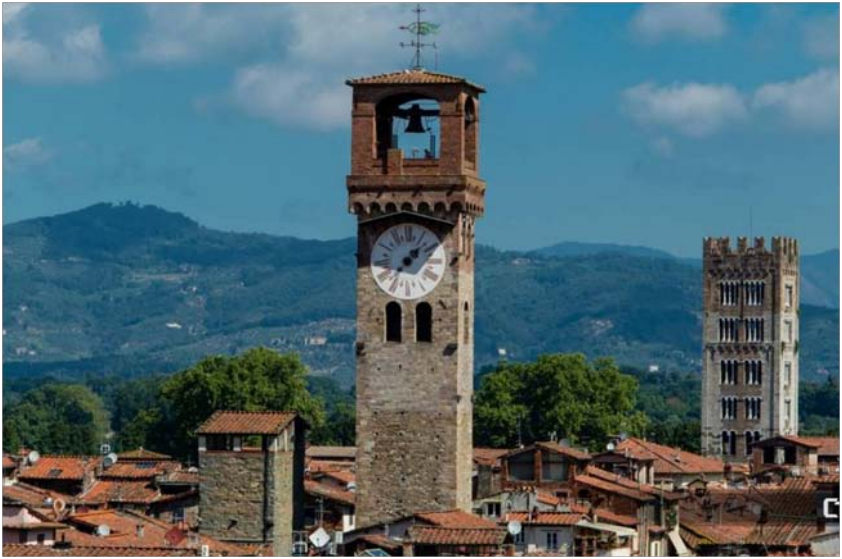

Figure 6. Skyline of the Lucca historic centre: the Clock Tower.

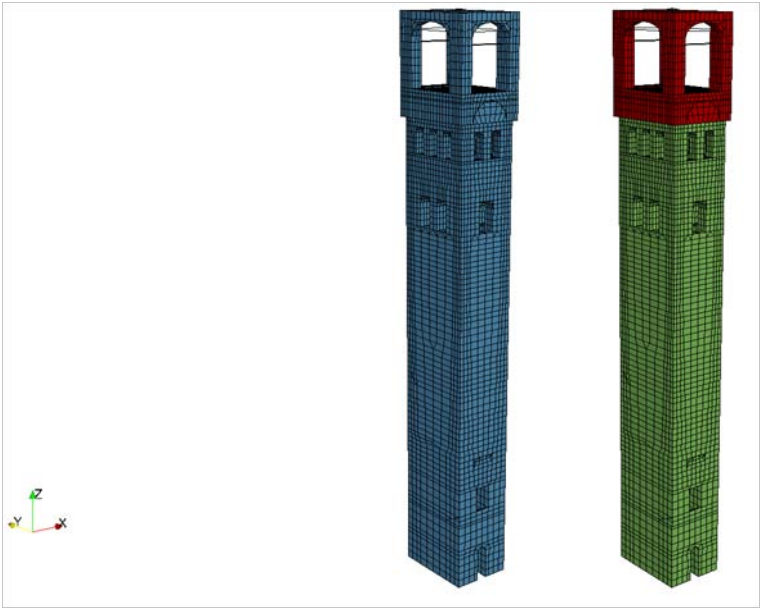

Figure 7. The Clock Tower, mesh 1 on the left, mesh 2 on the right.

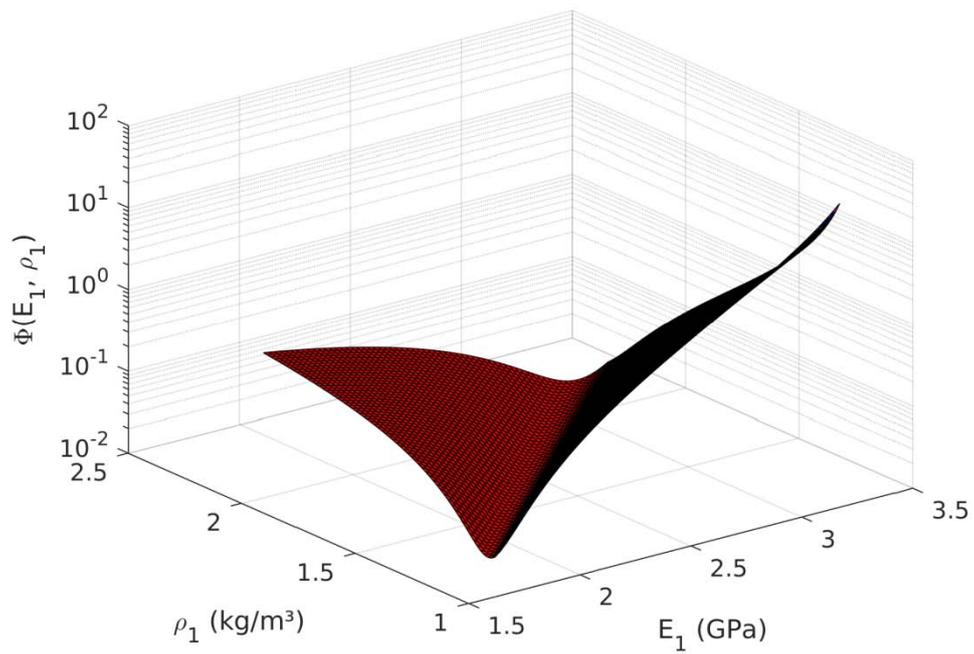

Figure 8. The Clock Tower, mesh 1: objective function $\Phi(E_1, \rho_1)$ vs. $E_1$ and $\rho_1$.

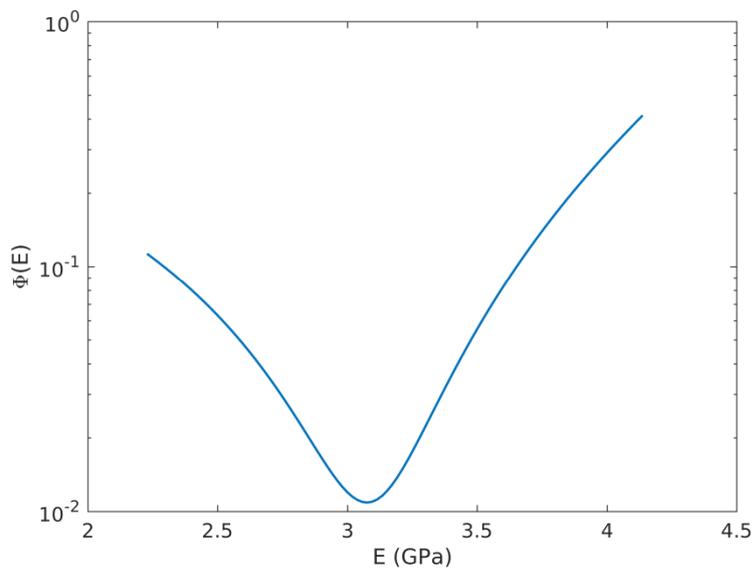

Figure 9. The Clock Tower, mesh 1: objective function $\Phi(E)$ vs. $E$.

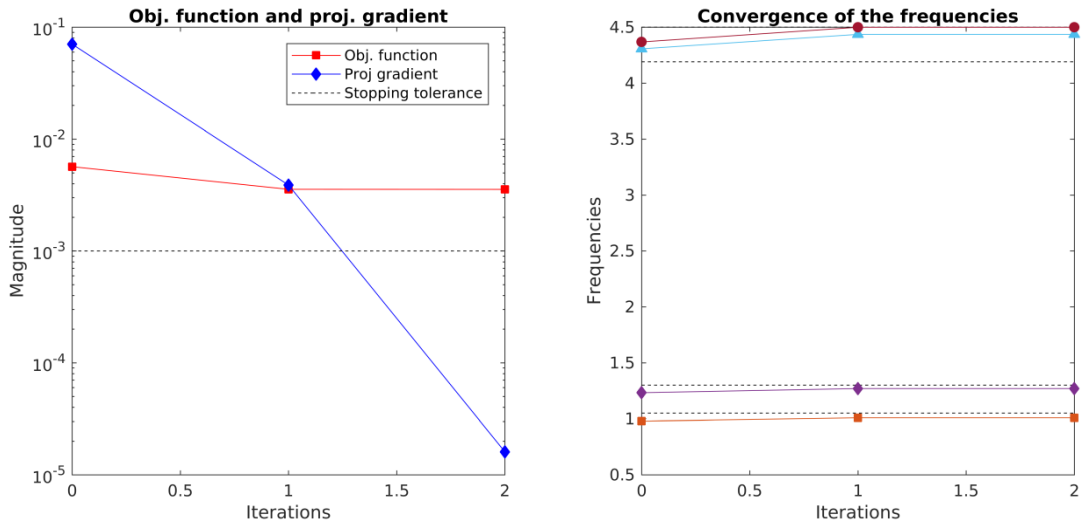

Figure 10. The Clock Tower, mesh 1: convergence history of the optimization algorithm.

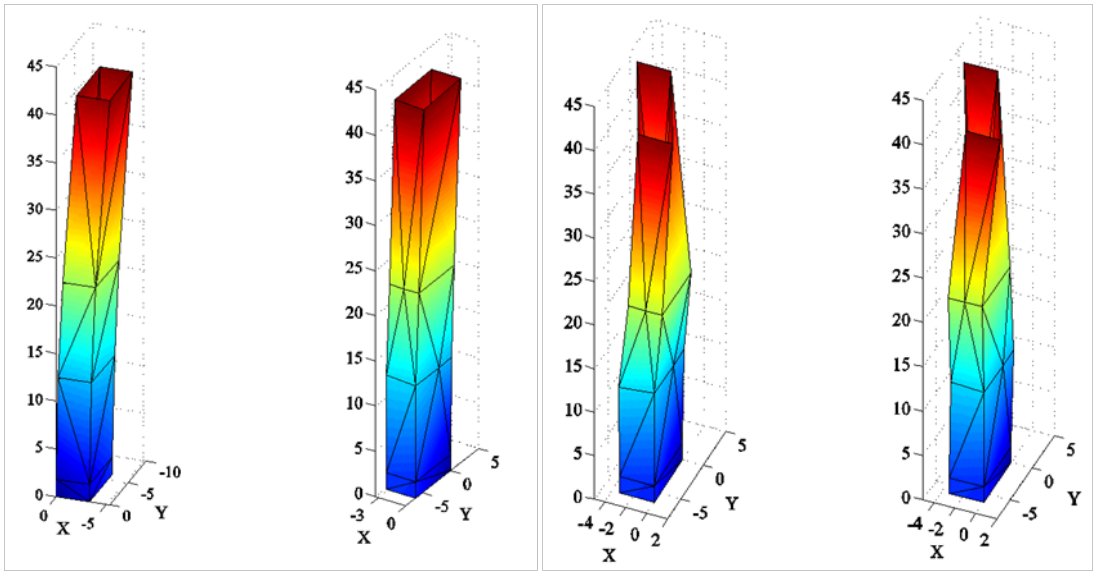

Figure 11. The Clock Tower, experimental mode shapes.

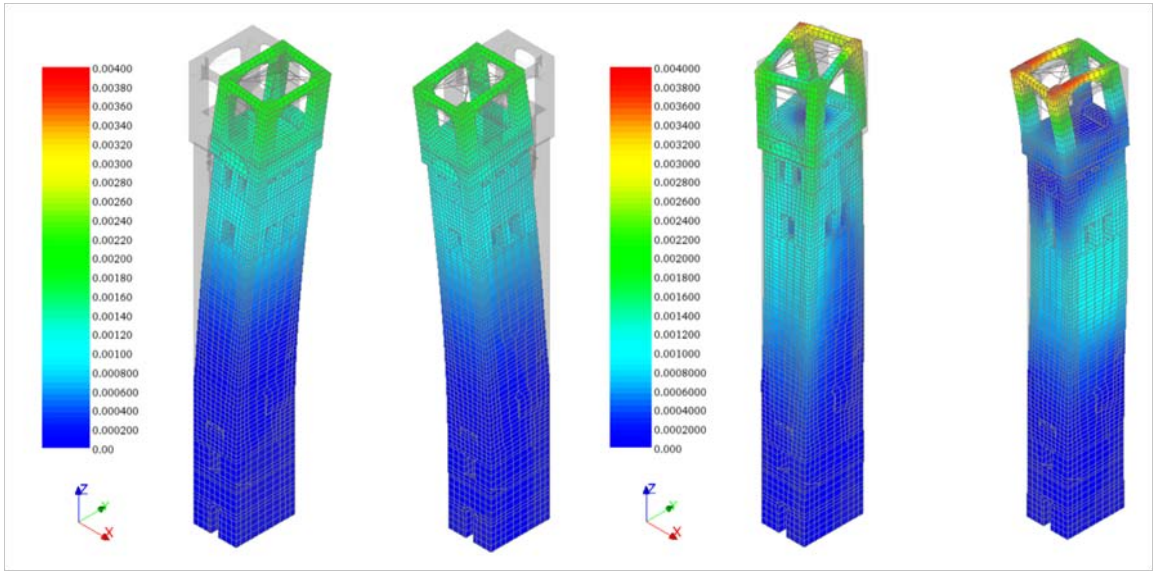

Figure 12. The Clock Tower, mesh 1: numerical mode shapes.

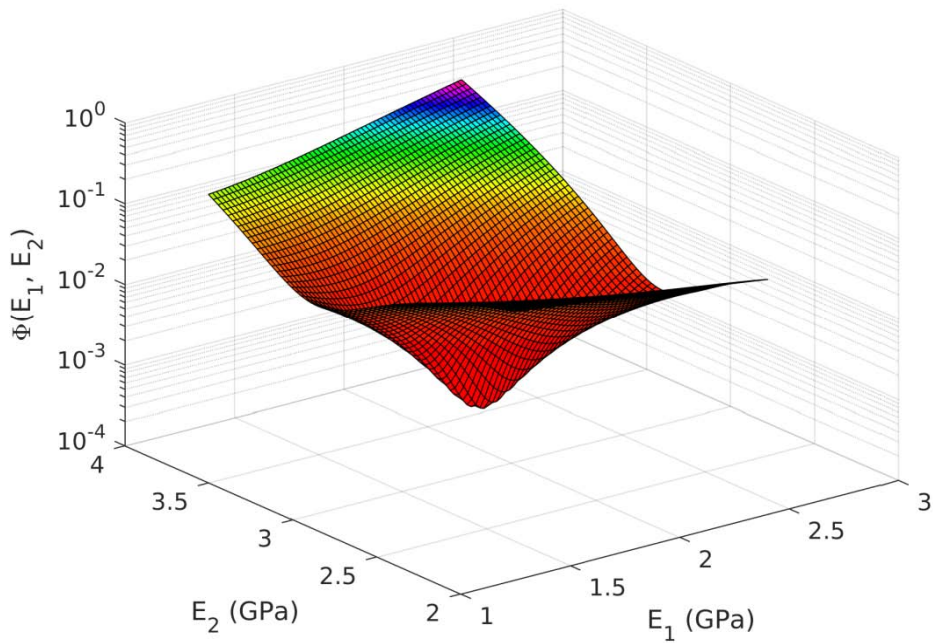

Figure 13. The Clock Tower, mesh 2: objective function $\Phi(E_1, E_2)$ vs. $E_1$ and $E_2$.

Final Draft of the paper https://doi.org/10.1061/(ASCE)CF.1943-5509.0001303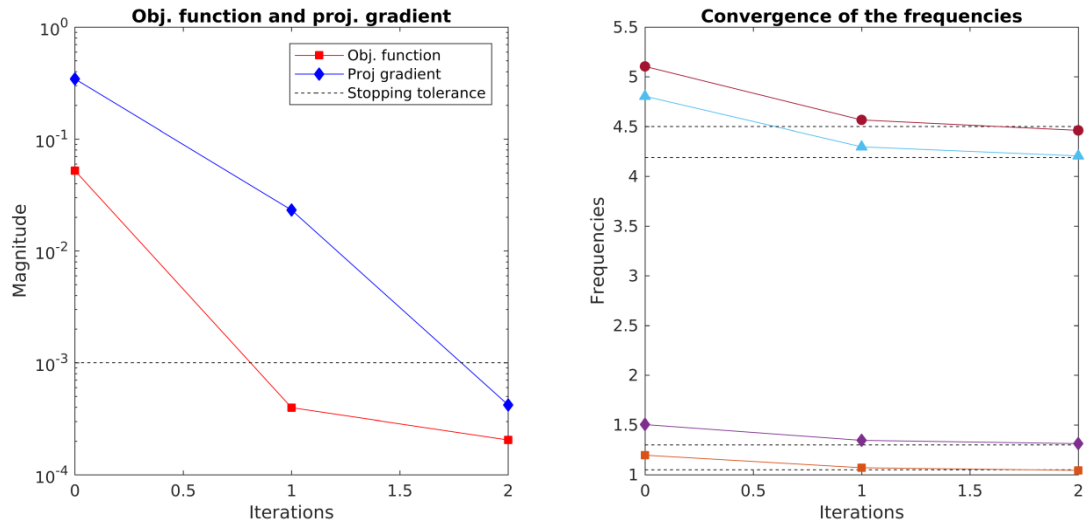

Figure 14. The Clock Tower, mesh 2: convergence history.

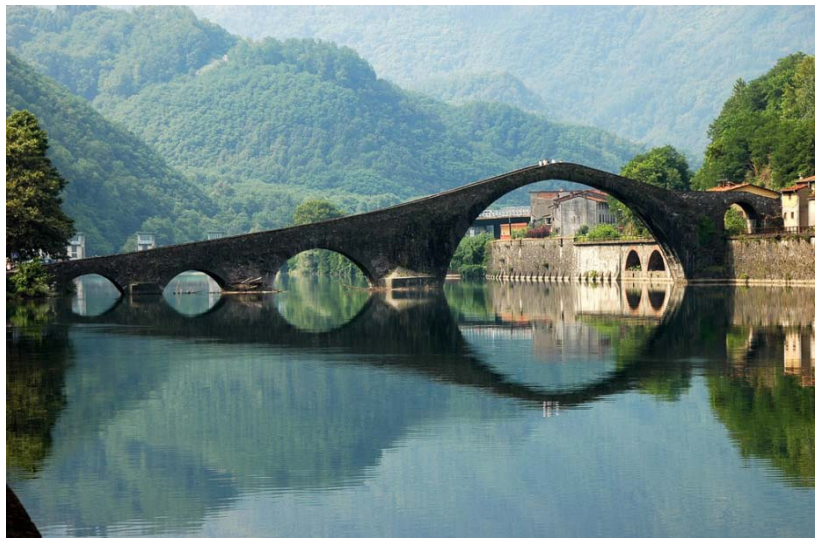

Figure 15. The Maddalena bridge in Borgo a Mozzano.

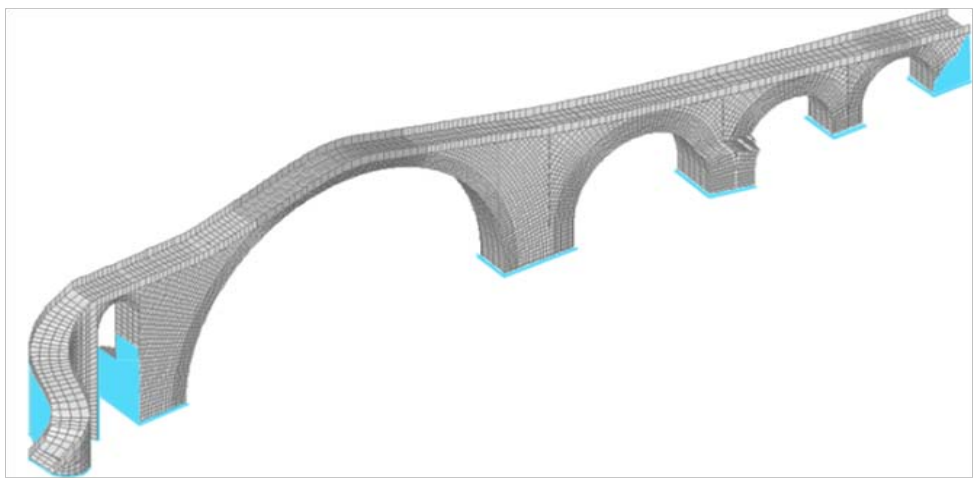

Figure 16. The Maddalena bridge: mesh 1.

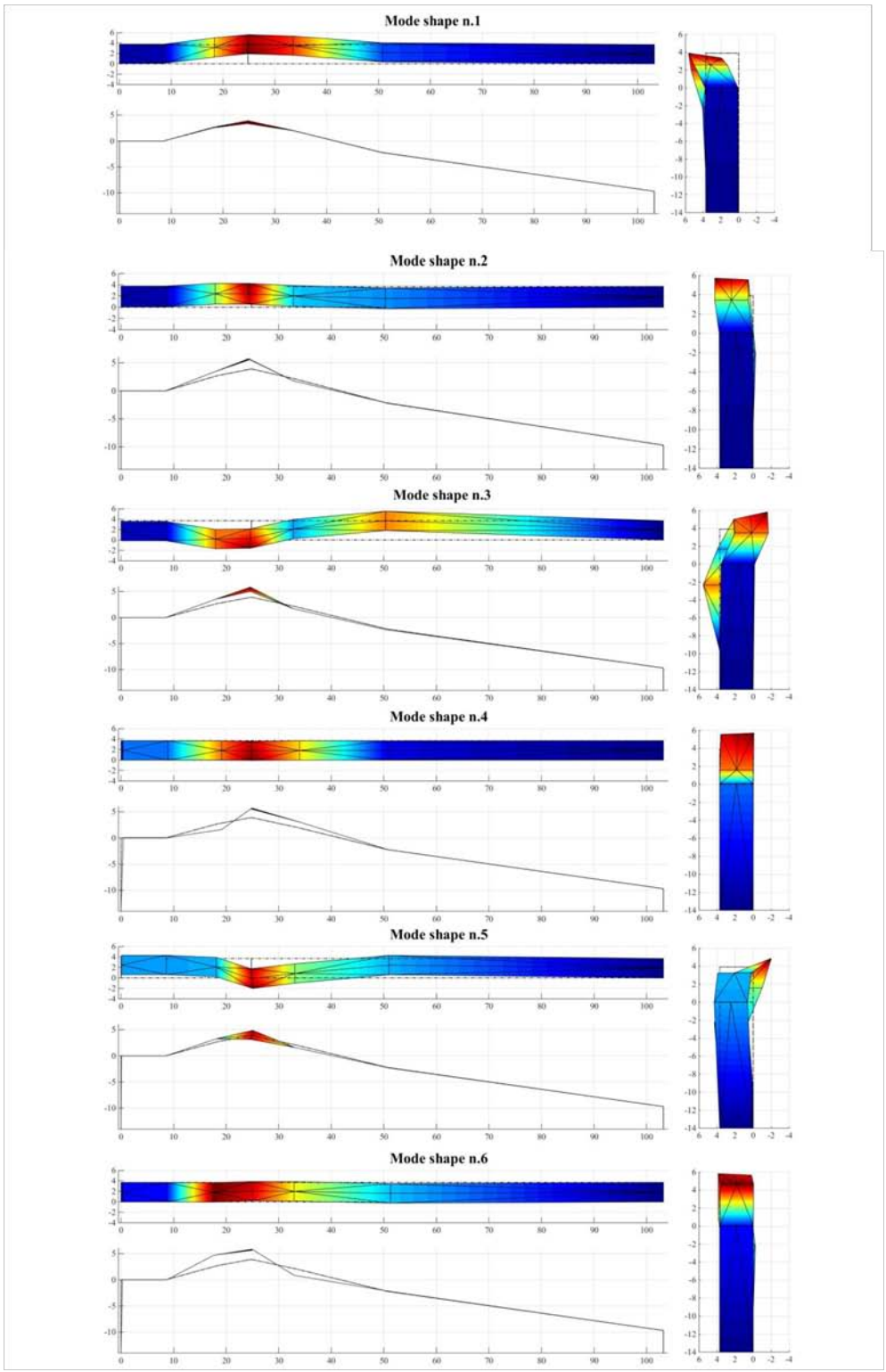

Figure 17. The Maddalena bridge: experimental mode shapes.

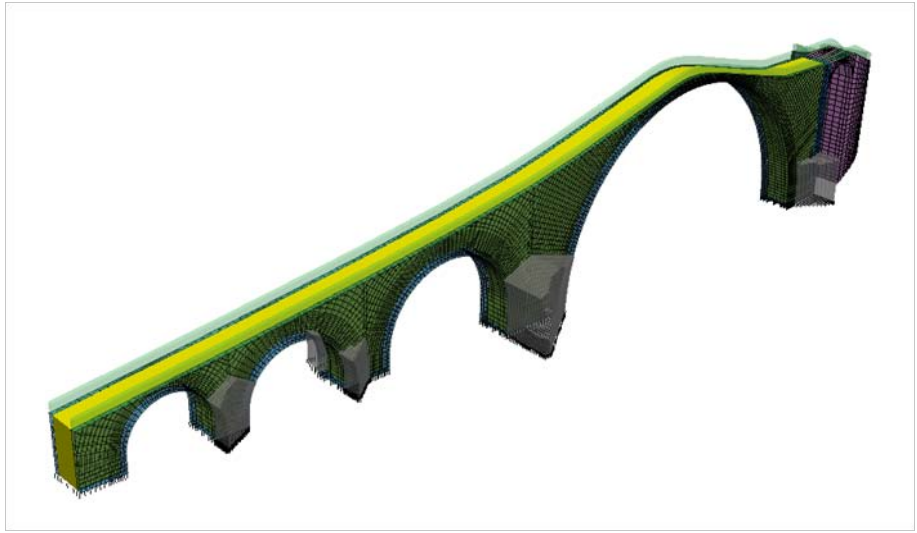

Figure 18. The Maddalena bridge: mesh 2.

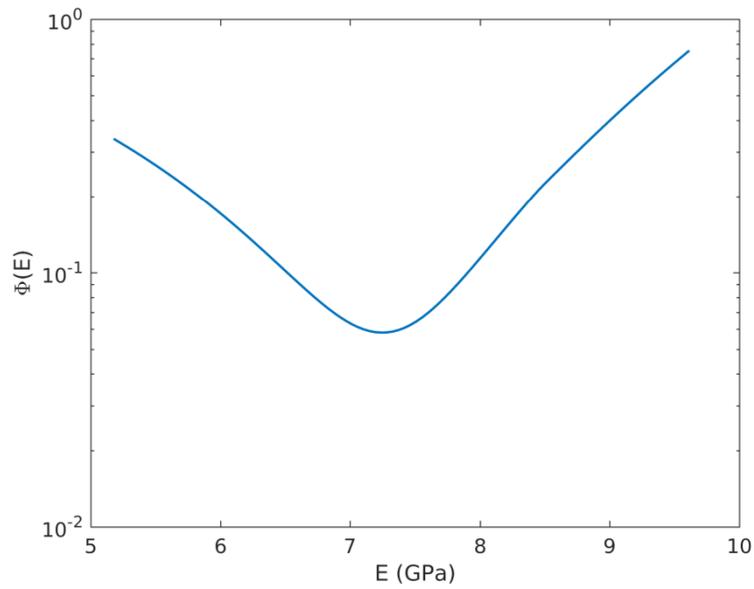

Figure 19. The Maddalena Bridge, mesh 1: objective function $\Phi(E)$ vs. $E$.

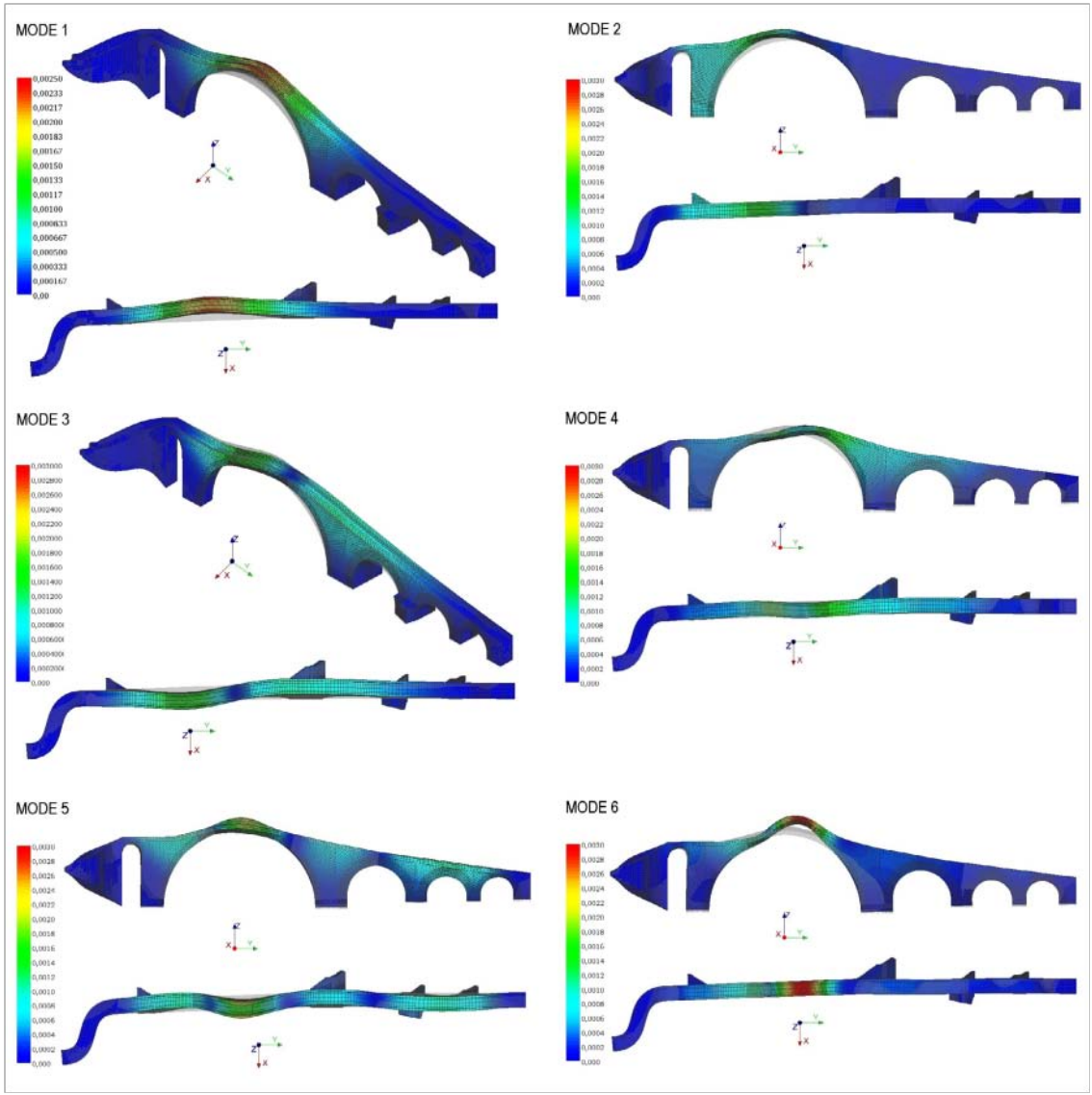

Figure 20. The Maddalena bridge: numerical mode shapes

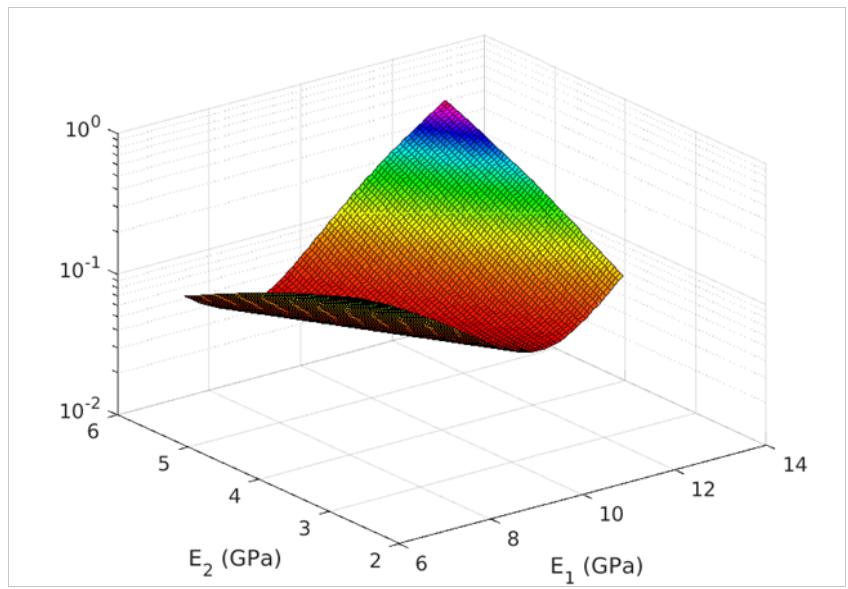

Figure 21. The Maddalena Bridge, mesh 2: objective function $\Phi(E_1, E_2, \rho_2)$ vs. $E_1$ and $E_2$.

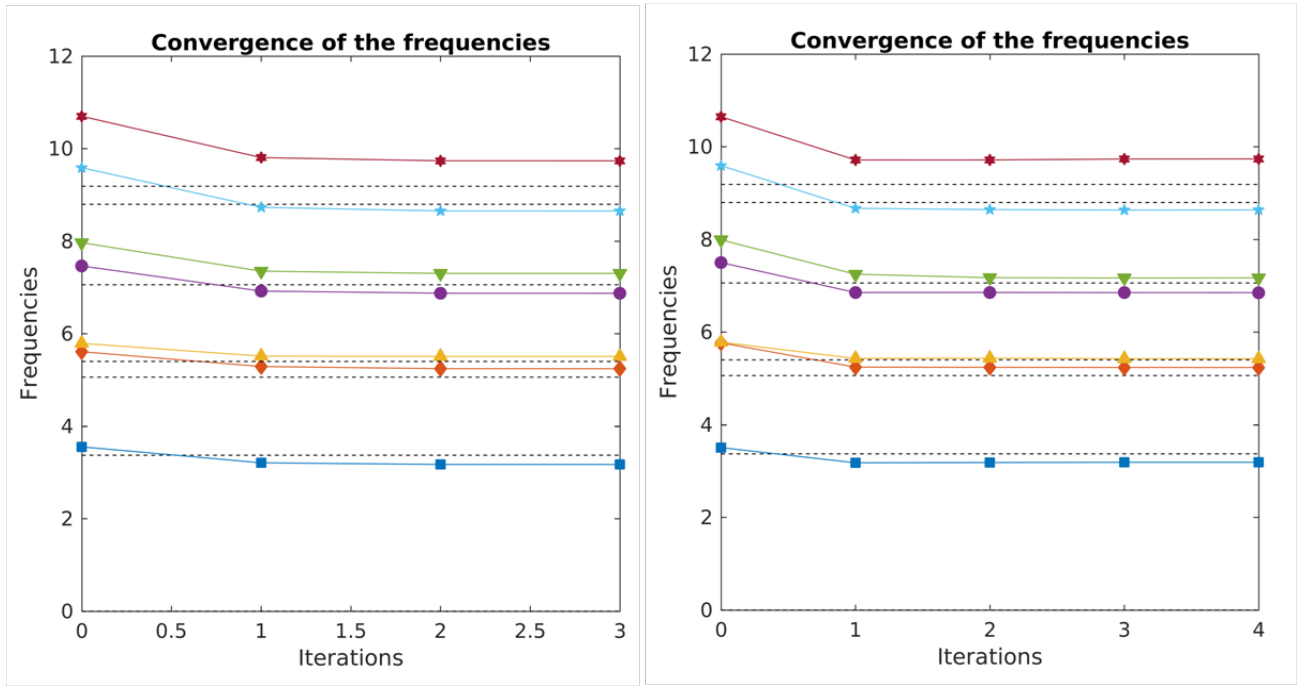

Figure 22. The Maddalena Bridge: convergence history for mesh 1 (on the left) and mesh 2.